# DS2SC-Agent: A Multi-Agent Automated Pipeline for Rapid Chiplet Model Generation

Yiwei Wu, Yifan Wu, Yunhao Xiong, Dengwei Zhao, Jiaxuan Shen
Jianfei Jiang, *Member, IEEE*, Guanghui He, *Member, IEEE*, Shikui Tu, *Member, IEEE*
and Yanan Sun, *Senior Member, IEEE*

*Abstract*—Constructing behavioral-level chiplet models (e.g., SystemC) is crucial for early-stage heterogeneous architecture exploration. Traditional manual modeling is notoriously time-consuming and error-prone. Recently, Large Language Models (LLMs) have demonstrated immense potential in automating hardware code generation. However, existing LLM-assisted design frameworks predominantly target highly structured or well-defined design specifications. In practical engineering scenarios, raw datasheets typically encompass lengthy, complex, and highly unstructured information. Consequently, reliable code generation directly from these raw datasheets suffers from severe challenges, including context vanishing and logical hallucinations.To overcome this critical bottleneck, this paper proposes DS2SC-Agent（Datasheet-to-SystemC-Agent）: the first end-to-end, fully automated generation pipeline that translates raw datasheets directly into SystemC chiplet models. This system establishes a highly efficient multi-agent collaborative framework. By decoupling the intricate modeling tasks, the proposed pipeline orchestrates a fully automated workflow encompassing unstructured long-document parsing, SystemC core code construction, testbench stimulus generation, and adaptive closed-loop debugging. We comprehensively evaluate the proposed framework on representative single-function chiplets across the analog, digital, and radio frequency (RF) domains—specifically, a Limiting Amplifier (LA), a Fast Fourier Transform (FFT) module, and a Power Amplifier (PA). The evaluation demonstrates that our pipeline seamlessly processes complex real-world datasheets to consistently generate functionally correct SystemC models. This provides a highly efficient and reliable paradigm for agile model library construction while drastically minimizing manual intervention.

## I. INTRODUCTION

In the post-Moore era, heterogeneous integration through chiplet technology has emerged as a primary driver for advanced semiconductor design, offering superior yield and high design reusability [1]-[3]. To perform efficient architecture exploration and hardware-software co-verification in the early stages of system integration, constructing precise and highly efficient behavioral reference models, such as SystemC models [4]-[6], is an indispensable step. However, the traditional workflow for constructing chiplet models relies heavily on domain experts [7]-[8]. Engineers are required to read through industrial-grade datasheets, which often spanning dozens of pages to extract critical interface specifications, timing constraints, and internal algorithmic logic, then subsequently translate this knowledge into SystemC code. This manual modeling process for diverse chiplets is inherently time-consuming, highly reliant on specialized expertise, and prone to human error, thus creating a severe bottleneck in agile hardware development.

Recently, Large Language Models have demonstrated immense potential in automating hardware code generation, significantly accelerating the design of Register Transfer Level (RTL) code such as Verilog [9]-[13]. Concurrently, a few pioneering studies have also explored the use of LLMs for SystemC code generation [14]. Despite the immense potential of LLMs, the vast majority of existing LLM-assisted hardware design frameworks evaluate model performance based on highly structured, simplified, or text-only design specifications [11].

In practical industrial engineering, however, raw datasheets are vastly different from clean specifications. A standard datasheet encompasses lengthy, complex, and highly unstructured information, incorporating natural language descriptions with parameter tables, timing diagrams, and architectural schematics. Directly feeding these raw, multimodal datasheets into an LLM inevitably leads to catastrophic failures. Single-pass (one-shot) generation approaches struggle with severe "context vanishing [15] ",where the model forgets initial interface constraints when generating internal logic and "logical hallucinations [16] ", making it nearly impossible to generate functionally correct SystemC code directly from industrial datasheets [17].

To overcome this critical gap between unstructured industrial datasheets and executable chiplet models, this paper proposes **DS2SC-Agent** (Datasheet-to-SystemC-Agent). This is the first end-to-end automated generation pipeline that translates raw datasheets directly into SystemC code, capable of supporting the agile modeling of chiplets across a diverse range of functions. DS2SC establishes a highly efficient multi-agent

This work was supported by the National Natural Science Foundation of China under Grant #92573106, and Shanghai 2024 "Science and Technology Innovation Action Plan" Fundamental Research Program in Integrated Circuit under Grant #24JD1400300. (Corresponding author: Yanan Sun)

Yiwei Wu, Yifan Wu, Yunhao Xiong, Jianfei Jiang, Guanghui He, and Yanan Sun are with the School of Integrated Circuits, Shanghai Jiao Tong University,Shanghai200240,China(email: {wwuyiwei, wu2389605167, xiongyhjdqx, jiangjianfei, guanghui.he, sunyanan}@sjtu.edu.cn).

Dengwei Zhao, Jiaxuan Shen, and Shikui Tu are with the School of Computer Science, Shanghai Jiao Tong University, Shanghai 200240, China (e-mail: {zdwccc, shenjiaxuan, tushikui} @sjtu.edu.cn).

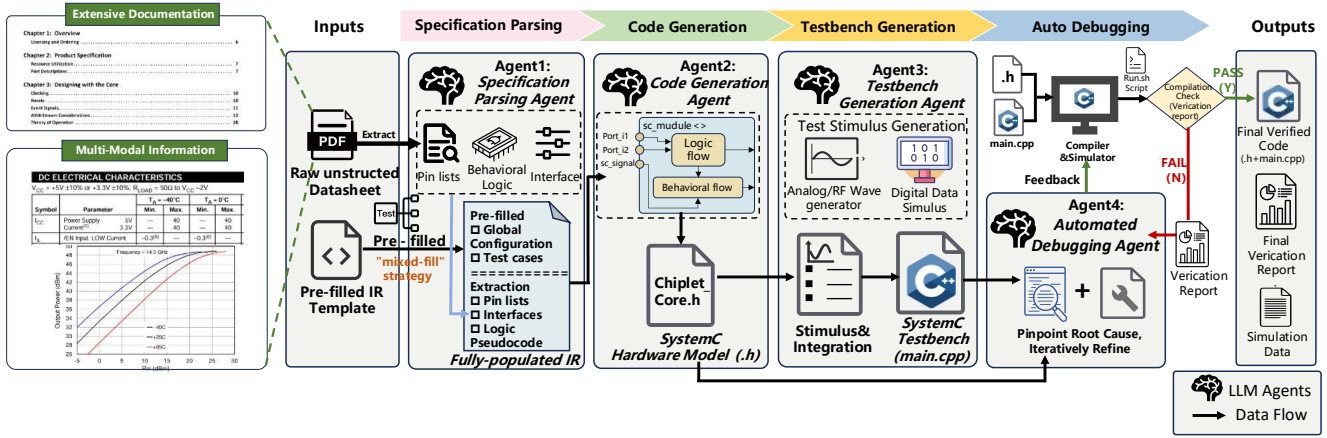

Fig .1: Overview of our proposed DS2SC-Agent system.

collaborative framework. By decoupling the intricate modeling tasks, the system orchestrates a fully automated workflow seamlessly driven by four specialized agents. Specifically, a **Specification Parsing Agent** first extracts critical parameters from the raw datasheet to generate a machine-friendly design Intermediate Representation (IR). Guided by this IR, a **Code Generation Agent** systematically constructs the hierarchical SystemC core code. Subsequently, a **Testbench Generation Agent** synthesizes a foundational testbench to drive the simulation, while an **Automated Debugging Agent** ensures functional correctness by providing adaptive, closed-loop iterative code refinement based on execution logs. The main contributions of this paper are summarized as follows:

(1) We propose **DS2SC-Agent**, the first end-to-end automated generation pipeline bridging the gap between complex, unstructured datasheets and SystemC models, significantly reducing the manual overhead in agile hardware design.

(2) We design a highly efficient 4-Agent collaborative architecture that effectively mitigates the context vanishing and logical hallucination issues commonly faced by LLMs when processing lengthy industrial documents.

(3) We conducted a comprehensive evaluation of the DS2SC framework across diverse domains, including digital, analog, and Radio-Frequency (RF) chiplets. Experimental results demonstrate that our closed-loop multi-agent system consistently yields functionally accurate SystemC models, where the generated code precisely reconstructs the complex physical behaviors and logical specifications of the target components. This validates the framework's high fidelity and reliability in constructing agile, heterogeneous chiplet model libraries.

The rest of this paper is organised as follows. Section II introduces the overall architecture of DS2SC-Agent. Section III details the key techniques and implementation mechanisms of the four collaborative agents. Section IV comprehensively evaluates our work through real-world chiplet modeling case studies. Finally, Section V concludes the paper and outlines future research directions.

## II. DS2SC-AGENT: AN OVERVIEW

Fig. 1 illustrates the overall architecture of the proposed DS2SC-Agent framework. The primary objective of DS2SC-Agent is to autonomously generate a fully functional and verified SystemC behavioral model based on two primary inputs: (1) an unstructured, raw PDF datasheet, and (2) a pre-filled Specification Intermediate Representation (Spec IR) that defines specific parameter configurations and simulation test cases. To achieve this, the framework eschews the traditional single-prompt Large Language Model approach in favor of a highly collaborative, four-stage multi-agent workflow. The system comprises four specialized agents: Specification Parsing, Code Generation, Testbench (TB) Generation, and Auto Debugging.

### A. Agent 1: Specification Parsing

The Specification Parsing Agent serves to standardize unstructured hardware documentation by taking a raw PDF datasheet and a pre-defined Spec IR template as dual inputs. Rather than generating specifications from scratch, this agent extracts critical hardware information such as pin lists, main features, and register maps to populate the predefined template. The resulting fully-populated Spec IR then serves as the single source of truth for all downstream agents.

### B. Agent 2: Code Generation (Chiplet Modeling)

Following the parsing phase, the Code Generation Agent utilizes the fully-populated Spec IR to construct the SystemC hardware model. The Code Generation Agent synthesizes a highly integrated and self-contained header file. This single file explicitly defines the sc_module architecture, ports, and internal signals, while simultaneously embedding the complete behavioral logic and algorithmic flows. This single-file design allows the subsequent testbench to directly #include the model for seamless compilation.

### C. Agent 3: Testbench Generation (System Integration)

The Testbench Generation Agent takes both the Spec IR and the previously generated core header as inputs to synthesize a SystemC testbench, outputting a main.cpp file. This agent parses the pre-configured simulation scenarios and specific test cases (e.g., specific sine wave inputs or predefined data

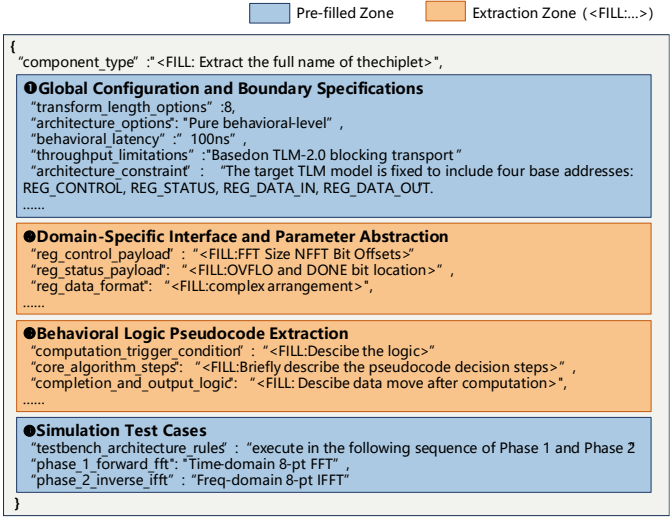

Fig. 2：The structure of the pre-defined Spec IR template based on the "mixed-fill" strategy

## A. Specification Parsing Agent

**Design and Cross-Domain Deconstruction of the Pre-defined JSON Template (Spec IR):** The Spec IR template serves as the cornerstone of the entire automated pipeline. Rather than a rudimentary enumeration of key-value pairs, the template's design encapsulates the domain expertise of front-end designers through a "mixed-fill" strategy. The template comprises two distinct data partitions: manually pre-filled constraints and reserved feature extraction anchors (uniformly denoted by the *<FILL:...>* tag). As illustrated in Fig. 2, the template is structurally divided into four core modules:

❶ **Global Configuration and Boundary Specifications (Pre-filled Zone):** Industrial datasheets typically encompass a vast array of configuration options and operating conditions. To mitigate hallucination during generation, we pre-populate this zone with deterministic system-level parameters. For digital chiplets, this includes fixed architectural parameters or behavioral-level latencies. For analog and RF chiplets, specific typical small-signal gains or baseline supply voltages are pre-configured. This pre-definition directly prunes the expansive design space, constraining the subsequently generated SystemC model to lock onto a specific, single-function configuration.

❷ **Domain-Specific Interface and Parameter Abstraction (Extraction Zone):** Given the substantial variations in interface specifications and key characteristics across different IP domains, the template employs domain-customized extraction strategies within this region, prompting the Agent to map complex information into the *<FILL:...>* placeholders: (1) For Digital Chiplets (e.g., FFT): The focus is on TLM memory map abstraction. Industrial interfaces often utilize complex bus protocols containing intricate low-level handshake signals. The template pre-defines standard register interfaces and constrains the Agent to accurately locate the bit widths and byte/bit offsets of critical signals from the datasheet, strictly mapping them into these standardized address spaces. (2) For Analog/RF Chiplets (e.g., LA, PA): The focus shifts to port definitions and core electrical specification abstraction. The Agent is required to extract and derive key features in the continuous-time domain from the mixed charts and parameter tables within the datasheet. These include port configuration rules (e.g., differential pins), calculation formulas for signal swings and static operating points, as well as non-linear transition points characterizing device behavior (such as saturation limits) and typical power consumption.

❸ **Behavioral Logic Pseudocode Extraction (Extraction Zone):** In this zone, the template utilizes the *<FILL:...>* tags to guide the Agent in extracting core dynamic behavioral rules. For digital logic, it primarily extracts initial reset logic and state machine transition conditions. For analog and RF logic, it focuses on extracting enable control sequences, linear amplification transfer functions, and non-linear mathematical expressions. This establishes a structured C/C++ style branch logic for downstream code generation, significantly mitigating logical inconsistencies during code generation by the LLM.

transform sequences) that are already populated within the Spec IR. The agent instantiates the generated SystemC module and translates these predefined scenarios into executable test stimuli and response-checking logic within the main.cpp.

## D. Agent 4: Auto Debugging

The Auto Debugging Agent orchestrates the closed-loop verification of the generated pipeline. The generated SystemC code and SystemC TB (*main.cpp*) are compiled and executed in a standard C++ simulation environment [6]. If a compilation or simulation failure occurs, an Error Log is generated. The Auto Debugging Agent parses this Error Log alongside the generated code and Spec IR to pinpoint the root cause, iteratively refining and correcting both the SystemC core model and testbench until test pass, yielding the final verified SystemC code.

**Extensibility and Prompt Optimization:** The DS2SC-Agent pipeline is designed with a highly decoupled architecture to ensure long-term adaptability. Rather than being tightly coupled to a specific foundational model, the system allows developers to upgrade or substitute the underlying LLM for any individual agent as more advanced, domain-specific models emerge. Additionally, to facilitate high-fidelity code generation and effectively manage context windows, every agent operates on highly specialized prompt templates specifically engineered for the designated hardware modeling sub-task.

## III. DS2SC-AGENT: THE FRAMEWORK PIPELINE

We first present the specification parsing from raw datasheets to populate the Spec IR template (Section III-A) and the SystemC code generation for chiplet modeling (Section III-B), both serving as the foundation of our automated pipeline. We then describe the testbench synthesis for system integration (Section III-C) and the adaptive closed-loop debugging (Section III-D) to iteratively refine the generated models and ensure functional correctness.

> **●Role Definition & Task Assignment**
> You are a senior SystemC modeling architect and front-end design expert. Your task is to extract key micro-architecture specs and electrical characteristics from the raw Datasheet to complete a provided "mixed-fill" JSON template
> ……
> **●Strict Boundary Constraints**
> Anti-tamper Principle: You are only allowed to modify the designated <FILL:...> placeholders. Modifying any pre-filled constraints, fixed numerical values, or JSON keys is strictly prohibited.
> **●Denoising**
> Noise Reduction: Ignore physical packaging, manufacturing data, and setup/hold timing. Abstraction: For digital interfaces, focus on TLM2.0 memory mapping (ignore low-level handshakes). For analog/RF, extract formulas for static operating points and transfer functions.
> ……
> **●Output format construction**
> Output purely valid JSON text. No Markdown tags, no conversational text, and no citations. Fill in "null" if a feature is absent. Do not hallucinate data.

Fig. 3: Prompt Instructions for Specification Parsing Agent.

❹ **Simulation Test Cases (Pre-filled Zone):** To provide a deterministic foundation for the downstream verification system (Agent 3), we pre-set core simulation test scenarios in this area. By pre-defining the expected test configurations and input stimulus characteristics, we ensure a strict alignment between the model generation and the verification objectives right at the specification parsing stage.

**Structured Design of the System Prompt for Agent1:** To maintain the LLM's high concentration and accuracy when processing lengthy, multimodal datasheets, this approach bypasses conventional natural language Q&A. Instead, we have designed a highly structured prompt framework for the Specification Parsing Agent, as illustrated in Fig. 3. By stratifying the instructions, this framework decouples the complex extraction task into four core dimensions, capable of adaptively accommodating the distinct requirements of digital and analog/RF chiplets:

❶ **Role Definition & Task Assignment:** The prompt initially configures the LLM with the profile of a "senior SystemC modeling architect" into the LLM. This contextual framing elicits the model's domain-specific priors associated with EDA, computer architecture, and circuit design. Simultaneously, it explicitly assigns the task of processing the "mixed-fill" template, which entails populating missing information while strictly preserving pre-existing constraints.

❷ **Strict Boundary Constraints:** To mitigate the issue of unintended parameter modification during generation, we enforce a critical immutability constraint within the prompt. The model is strictly restricted to operating exclusively within the workspaces demarcated by the <FILL:...> tags. Any alteration to the JSON key names or pre-filled parameters is immediately deemed a violation. This mechanism ensures structural integrity and data consistency.

❸ **Denoising:** Industrial datasheets contain extensive physical-layer specifications. The prompt instructs the extraction process to filter out irrelevant information such as packaging dimensions and manufacturing processes. More importantly, it provides a cross-domain abstraction paradigm: for digital chiplets, it forces the model to ignore low-level clock-cycle handshake details and focus entirely on TLM memory mapping; for analog chiplets, it directs the model to focus on parameter table lookups to extract continuous-time domain formulas and typical values (*Typ*).

❹ **Output Format:** At the output construction layer, the prompt enforces strict format sanitization. It demands output

> **●Role Definition & Task Isolation**
> You are a senior SystemC modeling architect. Your objective is to act as a structured hardware modeling engine to synthesize a pure Design Under Test (DUT) model from the provided Spec IR. Generating any Testbench code or toplevel sc_main functions is strictly prohibited.
> ……
> **●Architecture & Interface Synthesis**
> Model Definition: Inherit from the sc_module base class and instantiate standard communication ports based on the Spec IR interface features (e.g., TLM sockets for digital, continuous-time ports for analog/RF). Strictly use C++ STL containers and bitwise operations to construct internal registers according to the provided bit mappings.
> **●Event-Driven Behavioral Threading**
> Processing Thread: Translate the Spec IR pseudocode into core C++ processing threads (e.g., SC_THREAD). Explicitly utilize SystemC event mechanisms (e.g., sc_event.notify()) for state transitions and insert precise latency delays (wait(sc_time)) to emulate realistic hardware execution based on the performance specifications.
> ……
> **●Header-only Output Constraint**
> Output the hardware target module strictly as a single, self-contained C++ header file. Package all class declarations, inline algorithms, and required dependencies (e.g., <systemc.h>) together. Ensure structural clarity with concise comments.

Fig. 4: Prompt Instructions for Code Generation Agent.

in pure, valid JSON format, eliminating all Markdown tags and meaningless conversational text. Concurrently, it establishes a "null-if-missing" principle, to minimize data hallucination in cases of incomplete datasheet information.

*B. Code Generation Agent (Chiplet Modeling Agent)*

Upon receiving the fully-populated Spec IR from the Specification Parsing Agent, the Code Generation Agent acts as a structured hardware modeling engine. The primary objective of the Code Generation Agent is to transform highly abstract, machine-readable specifications into high-quality, compilable SystemC behavioral chiplet models. To streamline subsequent system integration and automated compilation chains, this agent is strictly constrained to a header-only design paradigm. This approach aggregates module declarations and complex behavioral logic into a single, self-contained .h file. To ensure that the generated SystemC code meets industrial standards and maintains high executability, we have designed a structured prompt framework for this agent across four dimensions, as illustrated in Fig. 4:

❶ **Role Definition & Task Isolation:** The prompt first assigns the persona of a "senior SystemC modeling architect" to the LLM. To prevent functional creep such as the unintended generation of extraneous test code, the instructions explicitly demarcate task boundaries. The input is restricted to the JSON-formatted Spec IR, and the output must be a pure hardware target module. The generation of any testbench classes or top-level simulation entry points, such as sc_main, is strictly prohibited.

❷ **High-Fidelity Architecture & Interface Synthesis:** At the structural level, the instructions constrain the generated architecture to strictly inherit from the sc_module base class. The agent is required to dynamically read the interface features from the Spec IR, such as TLM memory maps for digital chiplets or port definitions for analog/RF components, and then accurately instantiate the corresponding standard communication sockets or ports based on these extracted features.

❸ **Event-Driven Behavioral Threading:** For dynamic behavior, the prompt requires the model to translate the pseudocode within the Spec IR into efficient C++ processing threads (e.g., utilizing *SC_THREAD* or *SC_METHOD*). Two critical constraints are emphasized: first, an explicit trigger mechanism, where the model must implement thread awakening and state transitions via SystemC event mechanisms (e.g., *sc_event.notify()*) based on the initialization and trigger conditions defined in the IR; and second, hardware-level timing simulation, which requires the model to strictly retrieve latency parameters from the performance specifications and insert precise delays (e.g., wait(*sc_time*)) to reflect realistic hardware execution time.

Fig. 5: Prompt Instructions for Testbench Generation Agent.

❹**Header-only Output Constraint:** Regarding the output format, the prompt enforces a rigorous constraint requiring the LLM to package all class declarations, inline functions, and complex algorithmic derivations into a single C++ header file (e.g., *chiplet_core.h*). The agent must also automatically include all necessary dependencies (e.g., *<systemc.h>*, *<tlm.h>*) and provide structured, concise comments. This highly self-contained output mode eliminates complex multi-file compilation and linking dependencies, thereby significantly reducing interface coupling for downstream system integration (Agent 3) and substantially increasing the success rate of automated compilation.

*C. Testbench Generation Agent*

Following the construction of the hardware target module (*.h* file) by the Code Generation Agent, the Testbench Generation Agent initiates the preparation for closed-loop verification. Acting as an independent verification and system assembly generator, this agent automatically synthesizes a comprehensive black-box testbench [18]. The Testbench Generation Agent utilizes dual inputs: the Spec IR generated during the specification parsing phase (which includes predefined test cases and expected results) and the upstream-generated Design Under Test (DUT) header file. To facilitate fully automated compilation and simulation, the agent encapsulates the stimulus generator, the monitor/checker, and the top-level connection entry into a single, independently executable main.cpp file. To ensure precise interaction between the generated testbench and underlying models across various domains, we established a structured prompt framework comprising four dimensional constraints for this agent, as illustrated in Fig. 5; And the holistic code structures of the generated header-only model and the testbench assembly are visually summarized in Fig. 6.

❶**Role Definition & Black-Box Assembly:** The prompt

Fig. 6: Structural schematic of the generated artifacts: (a) the header-only chiplet model and (b) the independently executable testbench assembly.

assigns the LLM the persona of a senior SystemC/SystemC-AMS verification expert. The instructions explicitly mandate a black-box verification paradigm, treating the upstream-generated DUT as a closed system with unknown internal states. The DUT must be driven and observed exclusively through standard external interfaces (e.g., TLM sockets or Timed Data Flow (TDF) physical ports). This strictly isolates the design phase from the verification phase, preventing the homologous logic fallacy where identical conceptual errors propagate through both the design and test code.

❷**Stimulus Synthesis & Dynamic Synchronization:** For stimulus generation, the instructions constrain the model to strictly adhere to the pre-defined test scenarios within the Spec IR. For digital chiplets, the model is required to generate specific transaction-level test sequences and implement dynamic state polling with timeout mechanisms using SystemC *wait()* statements. For analog and continuous-time chiplets, the model must employ an internal state machine to sequentially switch waveform stimuli (e.g., small-signal, large-signal, and enable toggling) based on system timestamps, thereby ensuring precise synchronization of the simulation timeline.

❸**Simulation Data Logging & Verification Report Generation:** For runtime data recording, the prompt directs the agent to generate monitoring code that tracks I/O interactions. By extracting the current system simulation time (e.g., via *sc_time_stamp()*), it logs real-time, timestamped simulation data sequentially into a .csv file. For result evaluation, the model must automatically synthesize comparison logic or performance calculation formulas (e.g., for linear gain or voltage swing) based on the expected outcomes described in the Spec IR. Upon completion of the simulation, the generated testbench is designed to output a .txt verification report containing explicit PASS/FAIL indicators and detailed error analysis.

❹**Top-Level Execution Constraint:** In the final stage of output construction, the prompt enforces strict system-level interconnection constraints. The LLM must generate a standard *sc_main* top-level function at the end of the main.cpp

```
Role Definition & Dynamic Inputs
You are a SystemC expert, proficient in code review and bug fixing.
I will provide the following content:
SystemC Code: Including the header file (.h) and source file (.cpp or main.cpp).
Error Information: Terminal compilation output (for syntax errors) OR Simulation data .csv and
verification reports (for functional errors).
Your task is to repair the code based on the provided information and finally output the complete
corrected code.
Chain-of-Thought (CoT) Reasoning
Step 1: Analyze the root cause of the error based on the terminal error messages or the testbench
verification logs.
Step 2: Modify and correct the code based on the analysis results.
Step 3: Output the completely corrected C++ code block.
Mandatory Output Constraints
To enable the Python automation script to parse your output, you must encapsulate the corrected
code within a Markdown code block (```cpp).Inside the code block, the very first line MUST be a
special comment indicating the file name:
If repairing the header file, it must start with: // === FILE:chiplet_core.h ===
If repairing the main file, it must start with: // === FILE: main.cpp ===
If both files need to be repaired, please output two independent, complete code blocks, each
containing its respective file namecomment.Outputtingpartial code snippets is strictly prohibited.
You must output the fully modified, readyto-compile full-file content.
```

Fig. 7: Prompt instance for the Automated Debugging Agent, incorporating Chain-of-Thought reasoning and strict full-file output constraints.

file. This function is responsible for instantiating the DUT and testbench components, accurately binding all ports via signals or sockets, configuring optional VCD waveform tracing, and ultimately invoking *sc_start()* to trigger the underlying simulation engine. This structural constraint ensures a robust and automated transition from textual specifications to a fully executable simulation environment.

*D. Automated Debugging Agent*

Although the preceding agents employ strict structural constraints, Large Language Models inevitably generate syntax errors or deep logical flaws (e.g., hallucinations) when producing complex C++ and SystemC code. To achieve truly autonomous, end-to-end automation, we introduce the Automated Debugging Agent. Acting as the feedback optimizer of the entire framework, this agent dynamically switches debugging strategies based on the specific error types returned by the downstream toolchain, accomplishing self-repair of the code without human intervention.

The core workflow is decoupled into two parallel feedback loops: (1) Syntax Debugging Loop (Compilation Phase): When the generated .h or main.cpp files trigger a fatal error during the invocation of the underlying compiler (e.g., g++), the automation script immediately captures the stderr output stream from the terminal. The agent extracts these terminal logs which contain specific line numbers and type mismatch errors and correlates them with the erroneous source code to precisely fix syntax errors, such as missing semicolons, undeclared variables, or mismatched module connection ports. (2) Functional Debugging Loop (Runtime Phase): When the code successfully compiles but the simulation results fail to meet expectations (e.g., the verification report is marked as FAIL), the agent initiates deep logical debugging. At this stage, the automation script provides the agent with the runtime-recorded simulation data (.csv logs), the verification report (.txt), and the original Spec IR. The agent is configured to cross-reference the actual timing and numerical values exposed in the .csv with the expected behavior specified in the Spec IR. This comparison enables the localization of underlying functional defects, such as state machine transition deadlocks, non-linear formula calculation errors, or bit-width truncations. To maximize the agent's troubleshooting and repairing capabilities, we introduce the Chain-of-Thought (CoT) [19] reasoning and strict automated parsing constraints into the prompt, as illustrated in Fig. 7;

**Structured CoT Reasoning:** When faced with complex code tracing, directly prompting the model for repaired code often yields ungrounded outputs or non-deterministic modifications. Therefore, the instructions constrain the model to execute a strict three-step CoT before outputting code: Step 1 requires a deep analysis of the root cause of the error based on the terminal error or simulation log; Step 2 entails deducing a modification strategy for the code based on the identified root cause; and Step 3 is the actual generation of the corrected code. This reasoning-before-execution paradigm significantly enhances the repair success rate for complex logical errors.

**Deterministic Output Constraints:** To ensure that the peripheral Python automation script can stably extract the repaired code and overwrite the original files, the prompt enforces rigorous output boundary management. The LLM is required to encapsulate the repaired code within standard Markdown code blocks, and the first line of the block must contain a specified special file header comment (e.g., // === FILE: chiplet_core.h ===). Furthermore, it is strictly prohibited for the model to output only partial code snippets. It is forced to output the fully modified, ready-to-compile full-file content, effectively mitigating the instability associated with complex local code patching.

IV. EVALUATION

*A. Evaluation setting*

**Benchmark:** To evaluate the effectiveness and cross-domain generalization capability of our proposed multi-agent generation framework, we curated a benchmark suite comprising three highly representative, single-function industrial chiplets. Because fully automated SystemC and SystemC-AMS modeling directly from unstructured datasheets is a pioneering endeavor, direct baseline tools are currently unavailable for comparison. Consequently, our evaluation primarily focuses on the functional fidelity of the generated models with respect to the specifications in original datasheets. This benchmark suite spans three distinct physical domains:

(1) Fast Fourier Transform: Represents the pure digital domain, utilized to verify the framework's extraction of discrete-time algorithmic logic and the synthesis of standard TLM-2.0 memory-mapped interfaces.
(2) Limiting Amplifier: Represents the continuous-time analog domain, focusing on the framework's accuracy in modeling multi-stage physical behaviors (e.g., linear amplification and non-linear clamping) using SystemC-AMS Timed Data Flow nodes.
(3) Power Amplifier: Represents the RF domain, designed to test the framework's extraction of complex RF characteristics and non-linear transfer functions, as well as the framework's capability to fit specific performance curves.

**Input Modality Diversity:** Beyond the span of physical domains, to rigorously test the framework's robustness during the information extraction and specification parsing phases, the selected chiplets exhibit significant variances in datasheet length and the primary data modalities (mixed text, tables, and graphs) of the critical information, which allows for a comprehensive evaluation of the framework's ability to generalize across heterogeneous specification formats:

(1) FFT (96-page Comprehensive Guide): Sourced from the extensive 96-page Fast Fourier Transform v9.1 LogiCORE IP Product Guide [20]. This document amalgamates dense textual descriptions, timing diagrams, and exhaustive configuration tables. The primary challenge involves filtering extensive textual noise to precisely extract the four core register mappings required for control and status interactions.

(2) LA (6-page Dense Tables) [22]: Derived from a concise 6-page datasheet. Although it contains basic textual context, the core physical parameters are primarily embedded within dense tabular data (e.g., AC and DC Electrical Characteristics tables). This case requires the framework to possess cross-modal alignment capabilities, extracting precise static operating points from tables to guide the subsequent three-stage (linear, clamping, disabled) behavioral modeling.

(3) PA (18-page Curve Graphs) [21]: Sourced from an 18-page datasheet. Alongside regular textual specifications, the PA's critical non-linear RF behaviors rely heavily on performance curve graphs under various input/output conditions. This case focuses on the framework's mixed-modal understanding capability, evaluating the system's ability to extract and fit specific RF performance curves (e.g., AM-AM conversion characteristics) from graphical trends.

**Implementation Details:** In our experiments, we employed Gemini 3 Flash [23] via API as the core reasoning engine across all automated agents. The selection of this specific model is primarily attributed to its extended context window support, which is critical for processing and parsing extensive documents without information truncation (e.g., the 96-page FFT product guide). To optimize performance across different sub-tasks, we systematically configured the temperature hyperparameters throughout the pipeline. Specifically, the temperature for the Specification Parsing Agent (Agent 1) was set to 0.2 to ensure high determinism and strict adherence to the predefined JSON structural templates.

For the Code Generation and Testbench Generation Agents (Agents 2 and 3), the temperature was slightly elevated to 0.4, balancing strict architectural constraints with the realtime flexibility required for complex C++ logic synthesis. Meanwhile, the temperature for the Automated Debugging Agent (Agent 4) was configured to 0.3, enabling it to maintain rigorous Chain-of-Thought reasoning during root-cause analysis while retaining the adaptability needed to devise effective code patches. Finally, during the execution and validation phase, all generated hardware models (*.h*) and testbench assemblies (*main.cpp*) were compiled and simulated utilizing the standard GCC compiler, linked against the IEEE SystemC 2.3.3 core library and the SystemC-AMS 2.3.4 proof-of-concept simulator. To comprehensively validate the functional fidelity and generative robustness across these diverse physical domains, the following subsections present detailed case studies for each synthesized chiplet.

TABLE I
VERIFICATION RESULTS OF THE 8-POINT FFT/IFFT
CLOSED-LOOP TEST

| Index | Input $(Re, Im)$ | IFFT Output $(Re, Im)$ | Differential Error $(\Delta Re, \Delta Im)$ |
|---|---|---|---|
| 0 | (1,0) | (1,0) | (0,0) |
| 1 | (2,0) | (2, -2.7e$^{-16}$) | (0, 2.7 e$^{-16}$) |
| 2 | (3,0) | (3, 4.4 e$^{-16}$) | (0,4.4 e$^{-16}$) |
| 3 | (4,0) | (4, 3.8 e$^{-16}$) | (0,3.8 e$^{-16}$) |
| 4 | (5,0) | (5,0) | (0,0) |
| 5 | (6,0) | (6, -5 e$^{-17}$) | (0,5 e$^{-17}$) |
| 6 | (7,0) | (7, 4.4 e$^{-16}$) | (0,4.4 e$^{-16}$) |
| 7 | (8,0) | (8, -6.4e$^{-17}$) | (0,6.4 e$^{-17}$) |
| **[Verification Result: PASS]** | | | |

*B. Case Study 1: Digital Domain - FFT Chiplet*

In the digital domain case study, we evaluate the framework's capacity to extract core control logic from the complex 96-page datasheet [20] and synthesize a SystemC model with high algorithmic fidelity. To rigorously verify the mathematical accuracy of the generated FFT chiplet, the fully automated testbench executed a complete "time-domain to frequency-domain and back to time-domain" closed-loop test.

Specifically, the testbench fed a purely real 8-point monotonically increasing sequence (1 to 8) into the generated FFT model. After completing the data transaction via the TLM bus and dynamically polling the status register to accommodate the hardware computation delay, the testbench retrieved the frequency-domain results. Subsequently, it fed these results back into the module to trigger the second phase: the Inverse Fast Fourier Transform (IFFT).

Table I presents the comparison between the final IFFT reconstructed output (Actual) and the initial 8-point sequence (Expected). Because the generated C++ algorithmic code incorporated standard double-precision complex data types (*std::complex<double>*), the actual output sequence accurately reconstructed the initial input. As shown in the table, the differential computational error (*Diff*) between the expected and actual values is bounded within the order of 10$^{-16}$. This negligible discrepancy inherently aligns with the machine epsilon limit for double-precision floating-point arithmetic. All data points successfully met the criteria of the automated tolerance checks, yielding a [VERIFICATION RESULT: PASS]. This result indicates that our framework not only constructs standard TLM-2.0 memory-mapped interfaces but also achieves high fidelity for the underlying complex DSP algorithms.

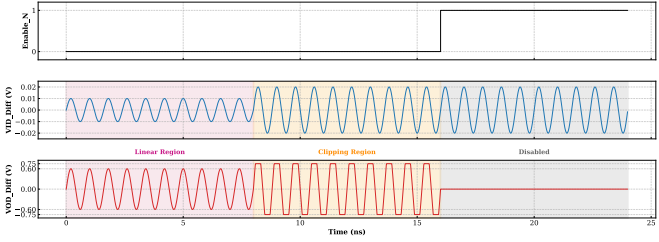

Fig. 8: SystemC-AMS transient simulation waveforms of the generated Limiting Amplifier (LA) model, illustrating the linear amplification, non-linear clamping, and disabled operational phases.

*C. Case Study 2: Continuous-Time Analog Domain - LA Chiplet*

We evaluated the framework's capability to extract precise static physical parameters from dense tabular data and construct a multi-stage SystemC-AMS Timed Data Flow model. For the Limiting Amplifier, the autonomously generated testbench injected a continuous-time sinusoidal stimulus with a progressively increasing amplitude, and dynamically toggled the "Enable" control pin specified in the datasheet. The simulation waveforms shown in Fig. 8 illustrate three distinct operational phases of the generated TDF model:

(1) **Linear Amplification Phase:** When the amplitude of the input sinusoidal wave resides within the small-signal range, the model exhibits ideal linear characteristics. The output voltage is linearly amplified in accordance with the voltage gain extracted from the AC electrical characteristics table.

(2) **Non-linear Clamping Phase:** When the amplitude of the input signal exceeds the predefined threshold, the limiting characteristics become active. The output waveform is clamped to the maximum output voltage swing limits specified in the DC characteristics table.

(3) **Disable/Enable Logic:** In the final stage of the simulation, the testbench triggers a state inversion on the /EN logic pin. The model transitions in response to this digital control signal; the TDF processing thread is logically bypassed, and the output returns to and maintains the predefined quiescent level. This validates the model's accurate response to digital control signals modulating continuous-time analog behaviors.

*D. Case Study 3: Radio Frequency Domain - PA Chiplet*

In the Radio Frequency domain case study, we evaluated the framework's capability to model the high-frequency non-linear behavior of a Power Amplifier, specifically focusing on the fitting accuracy of the critical Input-Output Power($P_{in}$ - $P_{out}$) relationship. Acknowledging the inherent limitations of Large Language Models in directly parsing complex graphical curves, this study adopted a pragmatic engineering strategy: fundamental physical parameters, such as small-signal gain ($G$) and saturation power ($P_{sat}$), were provided within the JSON prompt template, supplemented by several critical discrete data points extracted from the original $P_{in}$ - $P_{out}$ curve.

Given these foundational parameters and the targeted modeling approach, the framework successfully executed numerical parameter estimation. Based on the sparse sampled points, an appropriate non-linear smoothness factor (*s*) was derived to construct the SystemC-AMS behavioral code using the standard Rapp model to construct the SystemC-AMS behavioral code characterizing the PA's non-linear compression. To verify the fitting accuracy, the autonomously generated testbench executed an input power sweep simulation.

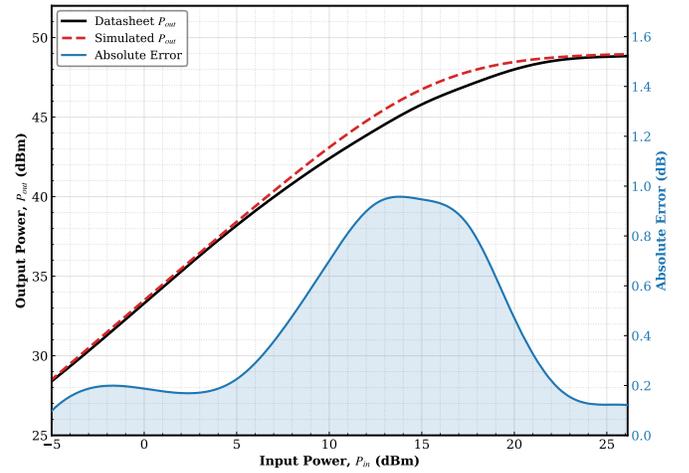

Fig. 9. Comparison of $P_{in}$-$P_{out}$ characteristics between the datasheet reference and the simulated results for the RF PA chiplet. The blue shaded area on the secondary axis represents the absolute fitting error, which is maintained within 1dB.

As illustrated in Fig. 9, the simulated $P_{out}$ exhibits high-fidelity fit with the ground-truth datasheet curve across the entire dynamic range, including the linear, compression, and deep saturation regions. The secondary axis (right) quantifies the absolute error, revealing that the maximum deviation remains strictly below 1dB even near the $P_{1dB}$ compression point. This conclusively demonstrates that the framework can effectively integrate a priori physical specifications with template directives, leveraging the system's algorithmic deduction capabilities to estimate missing parameters and reconstruct complex RF non-linear characteristics with high fidelity.

## V. CONCLUSION

This paper proposes a novel end-to-end multi-agent automated generation pipeline, DS2SC-Agent, designed to bridge the critical gap between unstructured, industrial-grade datasheets and executable SystemC chiplet models. To overcome the inherent limitations of large language models when processing long documents during single-pass generation such as context vanishing and logical hallucinations, we introduce a highly decoupled, four-agent collaborative architecture. By utilizing a "mixed-fill" Specification Intermediate Representation template, this pipeline systematically orchestrates automated workflows, including specification parsing, header-only core code construction, testbench synthesis, and adaptive closed-loop debugging, entirely without human intervention.

Comprehensive evaluations were conducted across diverse physical domains, encompassing a Fast Fourier Transform module in the digital domain, a Limiting Amplifier in the continuous-time analog domain, and a Power Amplifier in the radio frequency domain. These evaluations demonstrate that the proposed framework possesses robust cross-domain generalization capabilities and high simulation fidelity. Experimental results confirm that the generated models can accurately reconstruct complex algorithmic logic, multi-stage physical behaviors, and non-linear RF characteristics directly from raw, multimodal specifications. Ultimately, DS2SC-Agent establishes a highly reliable and efficient paradigm for the agile construction of chiplet model libraries, significantly reducing the substantial manual overhead traditionally required in early-stage heterogeneous architecture exploration.

Future research will focus on extending this multi-agent framework from single-function chiplet modeling to the automated generation and system-level verification of complex, multi-chiplet heterogeneous systems. Additionally, we plan to explore the integration of smaller, domain-specifically fine-tuned local models to further reduce inference latency and enhance data privacy in proprietary industrial applications.